\documentclass[preprint,proceedings]{rmaa}

\usepackage{paralist}

\usepackage{psfrag,color}
\usepackage{psfig}

\def\kmps{\hbox{$\km\s^{-1}\,$}}

\def\Msun{\hbox{$\rm\thinspace M_{\odot}$}}
\def\keV{{\rm\thinspace keV}}

\def\km{{\rm\thinspace km}}
\def\s{{\rm\thinspace s}}

\def\keV{{\rm\thinspace keV}}

\SetYear{2006}
\SetConfTitle{Triggering Relativistic Jets}

\title{Accretion Processes in AGN : The X-ray View} 

\author{
  Christopher S. Reynolds\altaffilmark{1}
\altaffiltext{1}{Dept. of Astronomy, Univ. of Maryland, College Park, MD~20742, USA}}

\shortauthor{Reynolds}
\shorttitle{Accretion Processes in AGN}

\fulladdresses{
\item Department of Astronomy, University of Maryland, College Park, MD~20742, USA. {\email{chris@astro.umd.edu}}.
}

\listofauthors{C. S. Reynolds}
\indexauthor{Reynolds, C.S.}

\abstract{We discuss constraints on the physics of the inner accretion
disk, as well as the properties of the black hole itself, that can be
derived by a detailed examination of the relativistically broadened
spectral features (especially the fluorescent iron line) in the
Seyfert galaxy MCG--6-30-15.  To begin with, we show that spectral
models which purport to eliminate the broad iron line in MCG--6-30-15
by invoking a moderately high ionization absorber are ruled out by
recent high-resolution spectra from the {\it Chandra} High Energy
Transmission Gratings.  We then discuss the comparison of {\it
XMM-Newton} data with accretion disk models.  The ``standard'' black
hole disk model of Novikov, Page and Thorne supplemented by the
so-called local corona assumption fails to produce sufficient
broadening; this indicates that the real accretion disk in
MCG--6-30-15 has significantly more centrally concentrated pattern of
X-ray irradiation that predicted by this model.  We discuss two
possible resolutions.  Firstly, the inner disk may be energized from
torques imposed by magnetic connections between the disk-proper and
either the plunging region or the rotating event horizon itself.
Secondly, X-ray emission from a high-latitude source (such as would be
the case of the X-ray source is actually the base of a jet) would be
gravitationally focused onto the central portions of the disk.  We
discuss how spectral variability may be used to examine these
possibilities and highlight the still outstanding mystery concerning
the anti-correlation between the iron line equivalent width and
relative normalization of the Compton reflection hump.  We end with a
few words about the exciting future of these studies heralded by {\it
Constellation-X} and {\it LISA}.}

\resumen{}

\begin{document}
\maketitle

\section{Introduction}
\label{sec:intro}

More than 40 years ago, it was suggested that the centres of galaxies
host supermassive black holes and, further, that accretion onto those
black holes powers active galactic nuclei (AGN; Salpeter 1964;
Zeldovich 1964; Lynden-Bell 1969).  Nowadays, the observational
evidence in support of this picture is substantial.  Proper motion
studies of the stars in the centralmost regions of the Milky Way
provide compelling evidence for the presence of a supermassive black
hole with a mass of about $3\times 10^6\Msun$ (Eckart \& Genzel 1997;
Ghez et al. 1998, 2000, 2003; Eckart et al. 2002; Sch\"odel et al.
2002).  The kinematics of rotating central gas disks in several nearby
low-luminosity AGN has also provided some of the most convincing
evidence for supermassive black holes (for example, M87: Ford et al.
1994, Harms et al. 1994; NGC~4258: Miyoshi et al. 1995; Greenhill et
al., 1995).  Finally, spectroscopic studies of stellar kinematics
reveal that almost all galaxies studied to date do indeed possess a
central supermassive black hole.  The very strong correlation between
the stellar velocity dispersion of a galaxy's bulge and the mass of
the black hole it hosts (Gebhardt et al. 2000; Ferrarese \& Merritt
2000) argues for an intimate link between supermassive black hole and
galaxy formation, a result of fundamental importance.

With the existence of supermassive black holes established, it is
clearly of interest to study them in detail.  While of crucial
importance for establishing the presence of supermassive black holes,
all of the kinematic studies mentioned above probe conditions and
physics at large distances from the black hole, $r>10^3r_{\mathrm g}$,
where $r_{\mathrm g}=GM/c^2$ and where $M$ is the mass of the black
hole.  However, the energetically dominant region of an AGN accretion
flow is very close to the central black hole, $r<20r_{\mathrm g}$, where
general relativistic effects become strong.  This is the region we
must consider if we are to truly understand these systems.  

Studies of the innermost regions of black hole accretion flows are
experiencing an unprecedented confluence of theoretical and
observational progress.  On the observational side, X-ray spectroscopy
of relativistically broadened and distorted emission lines (especially
the K-shell transitions of iron) are giving us a relatively clean
probe of the region within a few gravitational radii of several AGN
black holes (Tanaka at al. 1995; Fabian et al. 2000; Reynolds \& Nowak
2003).  These studies will be discussed in more detail in the
subsequent sections.  Theoretical progress has been equally dramatic.
With the realization that magnetohydrodynamic (MHD) turbulence driven
by the magneto-rotational instability is the basic driving mechanism
for accretion disks around compact objects (Balbus \& Hawley 1991),
the community is now constructing increasingly realistic,
first-principles simulations of black hole accretion disks.  For
example, current state-of-the-art simulations include the full
relativistic effects of a Kerr black hole (De~Villiers, Hawley \&
Krolik 2003), the relativistic mapping of disk emission to a distant
observer (Armitage \& Reynolds 2004), and accurate treatments of
radiation pressure (Turner 2004).  In a small number of years, these
simulations will be able to give first-principles predictions for
directly observable quantities, such as the energy spectrum, temporal
power spectrum and coherence function of the emitted radiation.

In this contribution, we discuss the constraints on the nature of the
black hole and inner accretion flow that have been derived through
X-ray iron line spectroscopy of the intensely studied Seyfert 1.2
galaxy MCG--6-30-15.  In particular, we discuss three main results.
Firstly, we show that the high-resolution spectrum of MCG--6-30-15
resulting from a half-megasecond observation by the {\it Chandra} High
Energy Transmission Gratings (HETG) gives strong support of the basic
broad iron line interpretation of the X-ray spectrum; this spectrum
allows us to rule out a model in which the 3-6\,keV spectrum curvature
(normally interpreted as the redshifted wing of the iron line) is due
to ionized absorption mimics a broad iron line.  Secondly, we show
that black hole spin is required to explain the breadth of the line; a
model based on a Schwarzschild black hole requires unphysical
distribution of emission across the disk surface.  Thirdly, we discuss
evidence for the effects of either strong light bending and interaction
between the disk and spinning black hole.

\section{The robustness of broad iron lines: new evidence from 
high-resolution spectroscopy}
\label{sec: robust}

\begin{figure*}
\begin{center}
\hbox{
\psfig{figure=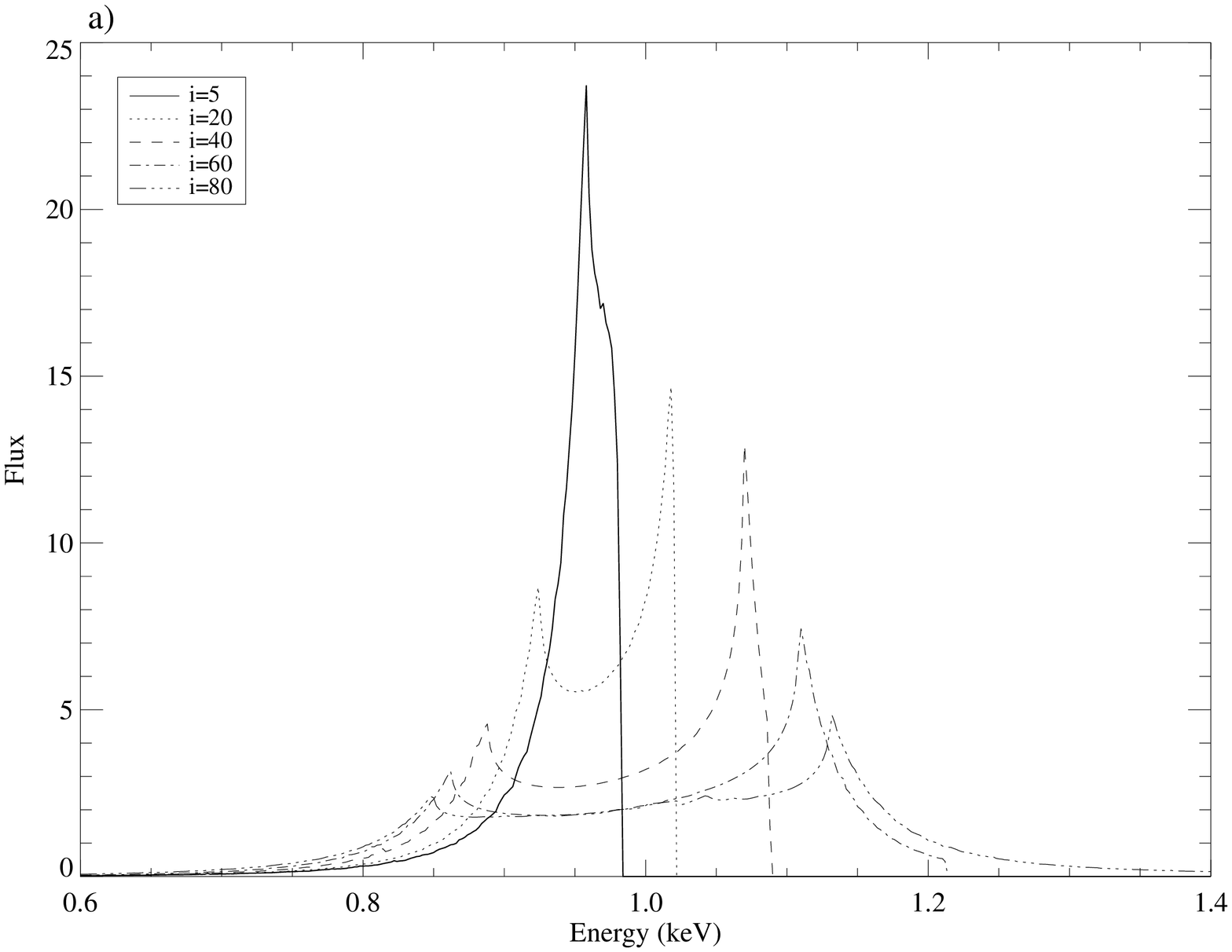,width=0.5\textwidth,angle=90}
\psfig{figure=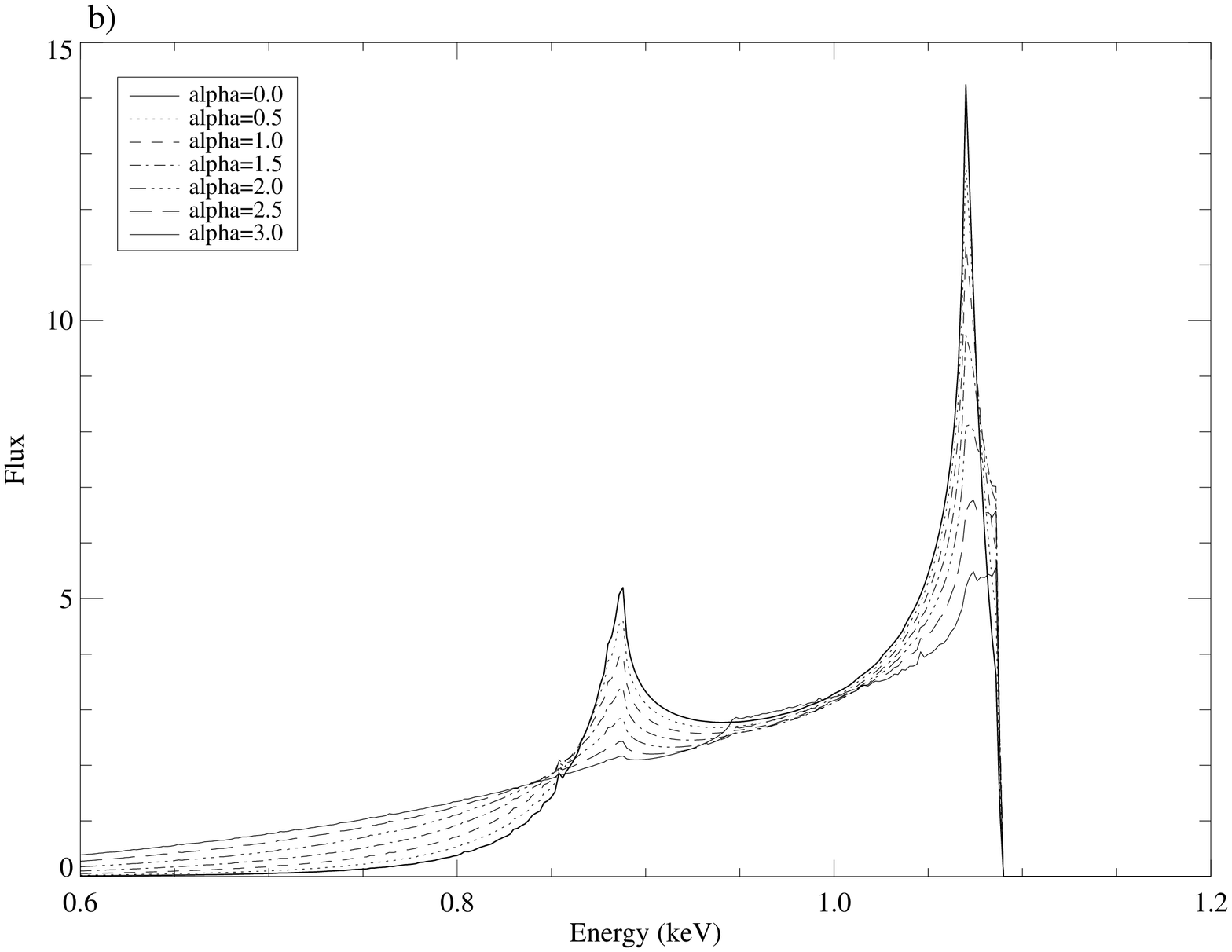,width=0.5\textwidth,angle=90}
}
\end{center}
\caption{Calculations of line profiles from a thin Keplerian disk
around a Kerr black hole.  In these examples, the rest-frame energy of
this hypothetical emission line is 1,\keV.  In all cases shown, the
range of emitting radii extends from the radius of marginal stability
to $50GM/c^2$.  {\it Left panel : }Variations of line profiles with
inclination angle for a radial emissivity proportional to $r^{-1.5}$
in the case of a rapidly-rotating black hole (dimensionless spin
parameter $a=0.998$).  {\it Right panel : }Variations of line profiles
with emissivity index $\alpha$ (such that the surface emissivity is
$r^{-\alpha}$) for a disk around a black hole with spin $a=0.5$ viewed
at an inclination angle $i=40^\circ$.  All computations have been
performed with the new {\tt kerrdisk} code (Brenneman \& Reynolds
2006).}
\end{figure*}

As we have already mentioned, the principal spectroscopic tool used to
date to study strong gravity is the characterization of the broad
iron-K$\alpha$ fluorescent emission line (see reviews by Fabian et al.
2000 and Reynolds \& Nowak 2003).  The essential physics underlying
this phenomenon is straightforward.  Moderate-to-high luminosity black
hole systems accrete via a radiatively-efficient disk.  Even in the
region close to the black hole, such a disk will (apart from a hot and
tenuous X-ray emitting corona) remain optically-thick,
geometrically-thin, almost Keplerian, and rather cold ($T<10^5$\,K for
AGN).  X-ray irradiation of the surface layers of the disk by the
corona will excite observable fluorescence lines, with iron-K$\alpha$
being most prominent due to the combination of its astrophysical
abundance and fluorescent yield.  This emission line is then subject
to extreme broadening and skewing due to the both the normal and
transverse Doppler effect (associated with the orbital velocity of the
disk) as well as the gravitational redshift of the black hole (see
Fig.~1).

Unfortunately, nature appears to conspire to ensure that we never have
an entirely ``clean'' view of a broad iron line.  In the AGN realm,
broad iron line signatures are usually observed in spectra that also
show clear signatures of absorption by circumnuclear photoionized
plasma (the so-called ``warm absorbers'').  This has led some authors
to suggest that the ``broad iron line feature'' is actually X-ray
continuum radiation that has been etched away by ionized absorption.
Within the context of these pure absorption models, strong curvature
is introduced below 7\,keV principally by L-shell absorption of
intermediate charge-states of iron.  Together with the sharp K-shell
edge at $>7\keV$, these models can produce a broad and asymmetric
feature that superficially resembles a broad iron line.  Such a model
was computed in detail for the case of MCG--6-30-15 by Kinkhabwala
(2003) and shown to provide an adequate fit to the time-averaged
350\,ks {\it XMM-Newton} data of Fabian et al. (2002) without any need
for relativistic effects.

\begin{figure*}[t]
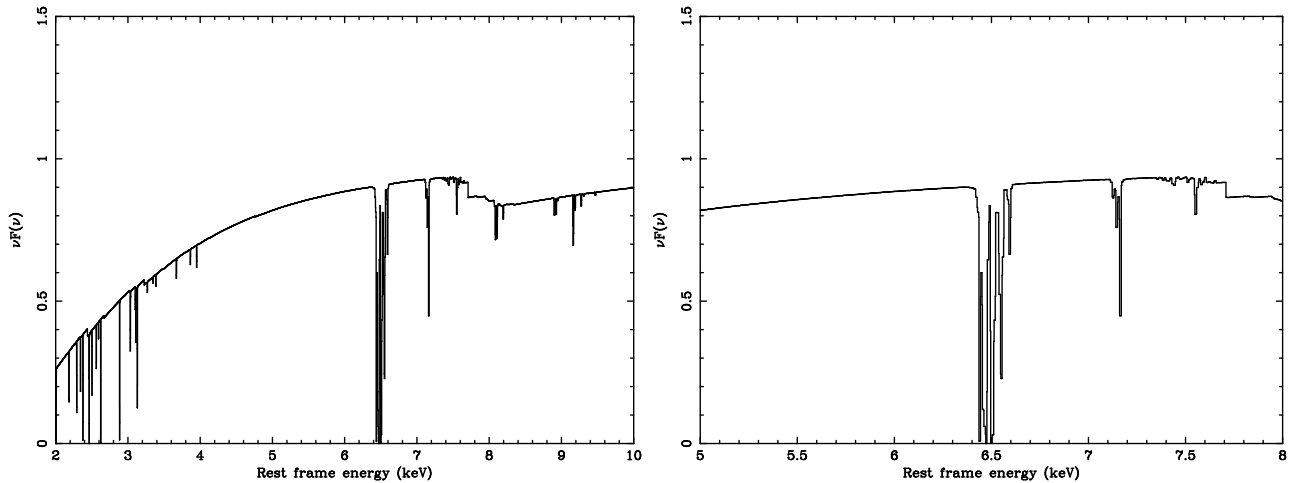

\begin{center}
\hbox{
\psfig{figure=f2a.ps,width=0.5\textwidth,angle=270}
\psfig{figure=f2b.ps,width=0.5\textwidth,angle=270}
}
\end{center}
\caption{Example of an ionized absorber model which will mimic a
broad iron line when observed at the moderate spectral resolution of
the {it XMM-Newton}/EPIC.  Parameters characterizing this absorber are
$\log\xi=2.2$ and $\log N_{\rm H}=22.6$ (cgs units).  The underlying
powerlaw has a photon index of $\Gamma=2$ and arbitrary normalization.
Note the deep absorption complex at approximately 6.5\,keV
corresponding to the resonant K$\alpha$ absorption lines of
intermediate ionization states of iron.  Models have been computed
using the XSTAR photoionization code.}
\end{figure*}

The Kinkhabwala ionized absorption model for MCG--6-30-15 makes a
generic prediction.  The intermediately ionized states of iron
(FeXVII--FeXXIII) producing the L-shell absorption noted above will
imprint a set of K-shell resonant absorption lines at 6.4--6.6\,keV
(see Fig.~2).  With the loss of the XRS on {\it Suzaku}, only the
Chandra/HETG has the spectral resolution required to search for and
characterize these lines.  With this goal, we\footnote{The Chandra
observation was made possible through the Guaranteed Time of
C.~R.~Canizares and A.~C.~Fabian.} obtained a 522\,ksec HETG
observation of MCG--6-30-15; the iron-band analysis of these data was
led by Andrew Young and has been reported by us in Young et
al. (2005).  This spectrum clearly shows a weak rest-frame narrow
iron-K$\alpha$ fluorescence line at 6.40\,keV, as well as the K-shell
resonant absorption lines of helium- and hydrogen-like iron (see
Fig.~3, left panel).  The highly ionized iron absorption lines are
blueshifted by $2000\kmps$ and hence can be attributed to a
photoionized wind, probably originating from the accretion disk
itself.  Most notable, however, are the {\it lack} of absorption lines
from intermediate ionization states of iron in this spectrum as
predicted by the Kinkhabwala model.

\begin{figure*}
\begin{center}
\hbox{
\psfig{figure=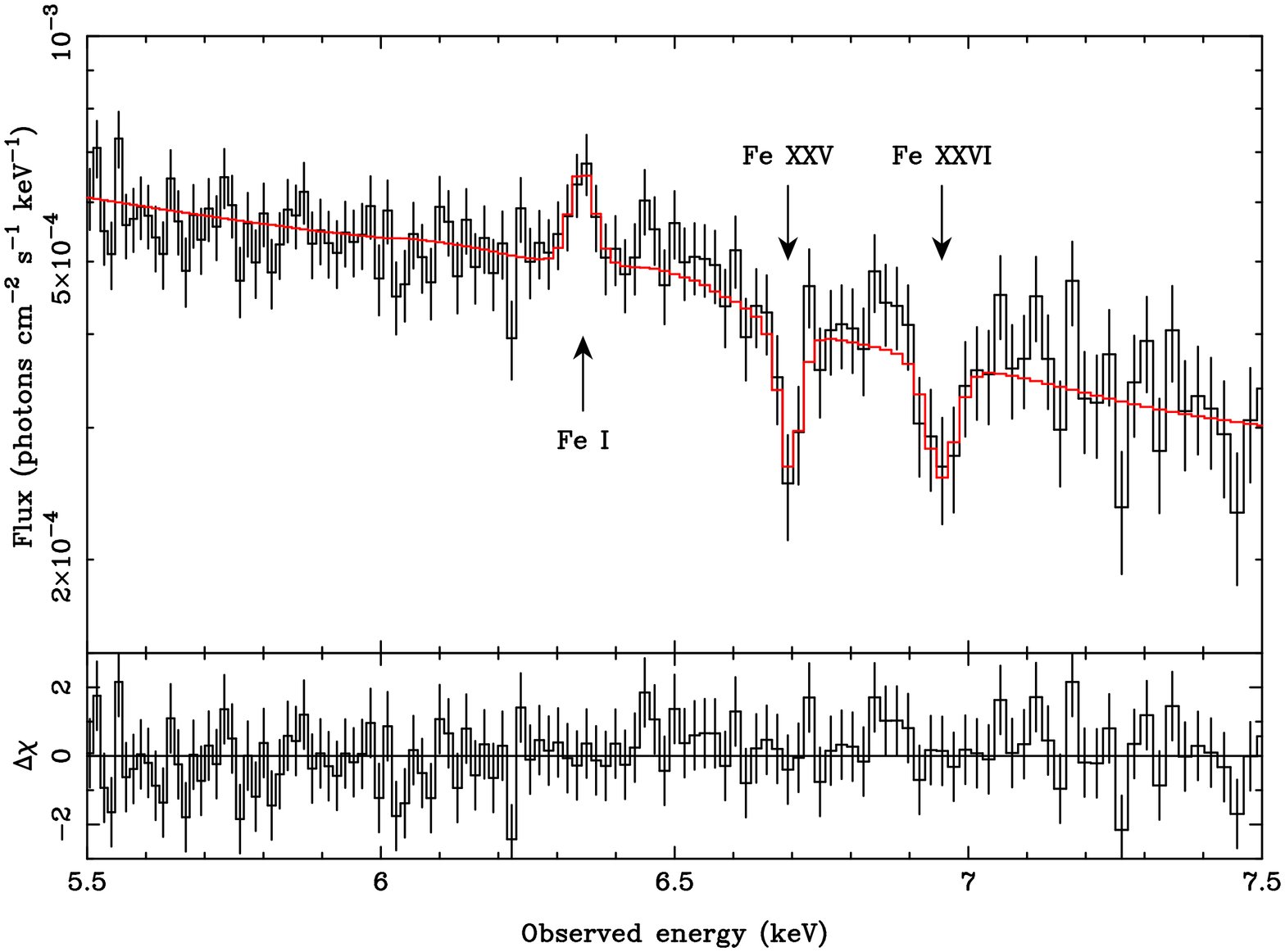,width=0.5\textwidth,angle=270}
\psfig{figure=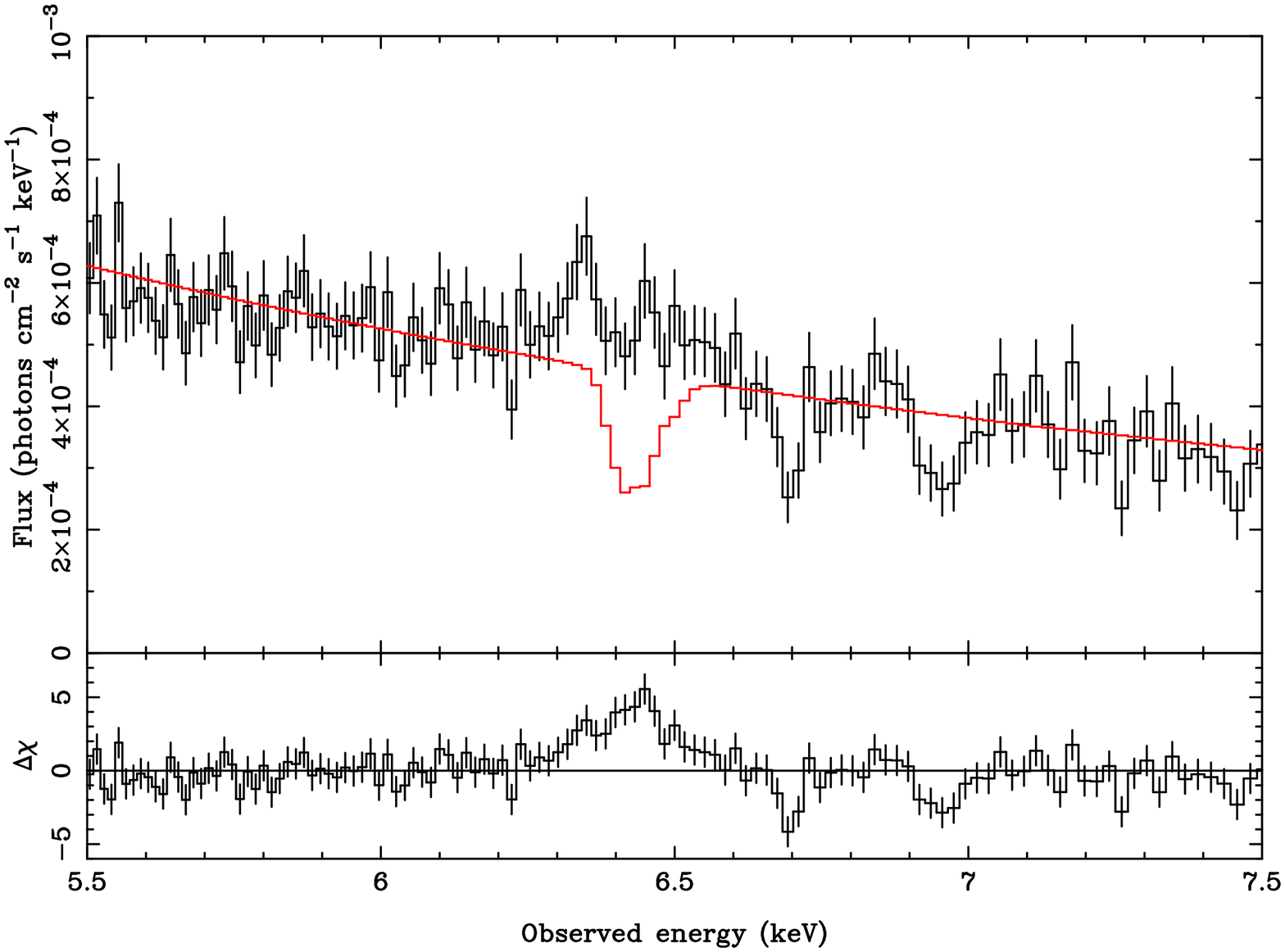,width=0.5\textwidth,angle=270}
}
\end{center}
\caption{High-resolution iron band spectrum of MCG--6-30-15 from the
{\it Chandra}/HETG.  {\it Left panel : }HETG spectrum overlaid with a
best-fitting model consisting of a power-law continuum, broad iron
line (too broad to be seen in this figure), narrow fluorescent line of
FeI (at rest with respect to the galaxy), and resonant absorption
lines of FeXXV and FeXXVI (blueshifted by $\sim 2000\kmps$ with
respect to the galaxy).  {\it Right panel : }HETG spectrum overlaid
with a model consisting of a power-law and a warm absorber with column
density and ionization state such that it eliminates the red-wing of
the iron line in the XMM-Newton/EPIC data.  Note the deep absorption
feature in the model at 6.5\,keV which is clearly absent in the data.
Figures from Young et al. (2005).}
\end{figure*}

We have performed a direct test of the ``broad iron line mimicking
warm absorber'' model for MCG--6-30-15 as follows.  Firstly, we
modeled the 3--10\,keV spectrum from the 350\,ks {\it XMM-Newton}/EPIC
observation with power-law subject to a warm absorber (modeled with
the XSTAR photoionization code) parameterized by an ionization
parameter ($\xi=L_{\rm ion}/nr^2$; where $L_{\rm ion}$ is the ionizing
luminosity of the source, $n$ is the electron number density of the
absorber and $r$ is the distance of the absorbing plasma from the
source) and a column density ($N_{\rm H}$); we also excluded the
6--8\,keV range from the fit as this is clearly affected by spectral
complexity including a narrow iron line.  Best fitting warm absorber
parameters are $\log\xi=2.2\pm 0.1$ and $\log N_{\rm H}=22.6\pm 0.1$
(cgs units).  We then fold the best fitting warm absorber through the
{\it Chandra}/HETG response matrix and compare it to our data (Fig.~3,
right panel).  The model displays a deep absorption feature at
6.5\,keV (rest-frame) which is clearly discrepant with our data.
Since these absorption feature originate from the {\it same ions} that
produces the broad iron line like spectral curvature, they are a
generic and unavoidable prediction of this model.  Thus, we conclude
that we can rule out a Kinkhabwala-like ionized absorption model,
further bolstering the claim that we are indeed seeing
relativistically broad iron line emission.

So, why do broad iron lines often appear in conjunction with ionized
absorption?  Many of the AGN displaying extremely broadened X-ray
reflection features are believed to be accreting at a significant
fraction of their Eddington limit.  These are precisely the same
systems that would be expected to possess a significant
radiatively-driven, photoionized wind; such a wind can be launched
from the accretion disk itself or be ablated by the radiation field
from the sides of the putative cold, dusty torus of AGN unification
schemes.  Indeed, analysis of warm absorber features in a sample of
AGN observed by the {\it Chandra}/HETG suggests that the ionized mass
outflow rates increase as with increasing Eddington ratio (McKernan,
Yaqoob \& Reynolds 2006).  Thus both broad iron lines and ionized
absorbers appear to be part of the phenomenology of moderately rapidly
accreting sources.

\section{Properties of the black hole and the innermost accretion flow}

Having established that the relativistic spectral broadening
hypothesis remains robust for the case of MCG--6-30-15, we will
proceed to discuss resulting constraints on the properties of the
inner accretion flow and the black hole itself.  We begin by
discussing constraints on the spin of the black hole.  We then proceed
to discuss the physics of the innermost regions of the accretion disk
and the X-ray source.  For the remainder of this section, we shall
discuss constraints derived from {\it XMM-Newton}/EPIC spectroscopy of
the broadened X-ray reflection features.

\subsection{Black hole spin}

There is a common misconception that rapidly spinning black holes {\it
invariably} produce broader and more highly redshifted emission lines
than slowly spinning black holes.  This stems from the fact that the
radius of marginal stability ($6GM/c^2$ for a non-rotating black hole)
for a prograde accretion disk pulls-in towards the horizon as the
spin-parameter of the black hole is increased.  Hence, the line
broadening will increase with black hole spin {\it if} the line
emission is always truncated at the radius of marginal stability (see
Fig.~4).  But it is important to realize that we can produce
arbitrarily redshifted and broadened emission lines from around even a
non-rotating black hole {\it if nature had the freedom to produce line
emission from any radius beyond the horizon} (Reynolds \& Begelman
1997).  This discouraging fact has led some authors to conclude that
current iron line profiles contain essentially no information on the
black hole spin (Dovciak, Karas \& Yaqoob 2004).

\begin{figure}
\begin{center}
\psfig{figure=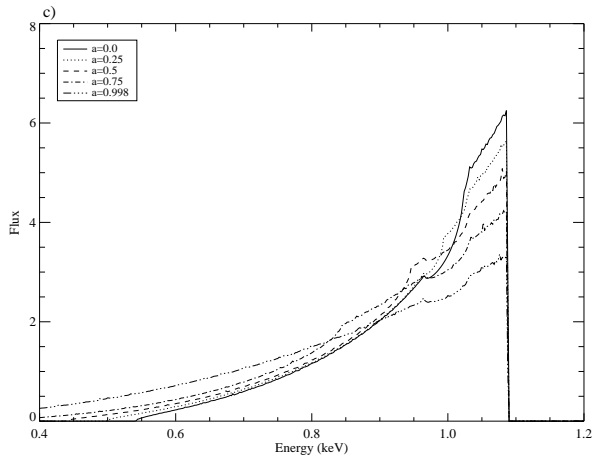,width=0.5\textwidth,angle=90}
\end{center}
\caption{As for Fig.~1, except showing variations of line profile as a
function of black hole spin for an inclination of $i=40^\circ$ and an
emissivity index of $\alpha=3$ under the assumption that the line
emissivity is truncated at the radius of marginal stability.  All
computations have been performed with the new {\tt kerrdisk} code
(Brenneman \& Reynolds 2006).}
\end{figure}

This would be an overly bleak assessment of our ability to constrain
black hole spin.  Even the application of some rather weak (i.e.,
general) astrophysical constraints can impose an inner limit on the
radii at which spectral features can be produced.  In order to produce
any significant iron emission line from the region within the radius
of marginal stability (which we shall refer to as the plunging
region), the disk in this region must be optically-thick, not too
highly ionized (i.e., a significant fraction of the iron cannot be
fully ionized), and illuminated by the hard X-ray continuum.  While
much work remains to be done on the physical state of matter in the
plunging region, it is challenging to construct a model for a disk
around a non-rotating black hole in which there are appreciable
spectral features produced by matter inside of $4.5-5GM/c^2$ (Reynolds
\& Begelman 1997).  If we require an emitting radius less than this
when fitting a non-rotating black hole model to a particular dataset,
we can claim to have found good evidence for a spinning black hole.

This is exactly the situation we find when attempting to fit the {\it
  XMM-Newton} data for the Seyfert-1 galaxy MCG--6-30-15.  Brenneman
  \& Reynolds (2006) perform a detailed re-analysis of the 350\,ks
  EPIC data (originally presented by Fabian et al. 2002) with a
  spectral model including a multi-zone dusty warm absorber and a
  relativistically-blurred reflection spectrum from an ionized
  accretion disk.  They find that, if one {\it imposes} a
  Schwarzschild geometry, any adequate fit requires most of the iron
  line emission to occur at a disk radius of $3GM/c^2$.  This is
  extremely deep within the plunging radius (i.e., the region within
  the radius of marginal stability at $6GM/c^2$), with radial
  velocities of the order of half the speed of light expected at that
  location.  Continuity of baryon number demands that the plasma at
  this location be extremely tenuous.  Thus, the irradiating X-rays
  required to produce the observed X-ray reflection signatures would
  photoionize this material to an extremely high degree, rendering it
  incapable of imprinting any atomic features on the spectrum.  Thus,
  we conclude that a Schwarschild black hole does not allow a
  physically viable model to be constructed for this spectrum.

In principle, the contribution of the plunging region to the broad
iron line profile can be quantified using a detailed model of X-ray
reflection from the plunging region which would need to include the
effects of finite optical depth and photoionization (Reynolds \&
Begelman 1997).  Significantly more theoretical work is required for
these models to stand a chance of describing reality.  However, one
can illustrate the potential power of broad iron lines to determine
black hole spin by the returning to the assumption that there
are no X-ray reflection features from the flow within the radius of
marginal stability.  As shown by Laura Brenneman elsewhere in this
volume, applying spectrum models employing this assumption to
MCG--6-30-15 suggests that the black hole is rapidly rotating,
$a>0.93$. 

\subsection{Confrontation between simple accretion theory and X-ray spectral data}

Even before the observational evidence for black hole accretion disks
became compelling, the basic theory of such disks had been extensively
developed.  Building upon the non-relativistic theory of Shakura \&
Sunyaev (1973), Novikov \& Thorne (1974) and Page \& Thorne (1974)
developed the ``standard'' model of a geometrically-thin,
radiatively-efficient, steady-state, viscous accretion disk around an
isolated Kerr black hole.  In addition to the assumptions already
listed, it is assumed that the viscous torque operating within the
disk becomes zero at the radius of marginal stability, $r=r_{\rm ms}$.
Physically, this was justified by assuming that the accretion flow
would pass through a sonic point close to $r=r_{\rm ms}$ and hence
flow ballistically (i.e., ``plunge'') into the black hole.  The
zero-torque condition at the radius of marginal stability leads to a
predicted dissipation profile which peaks at $r\approx 1.5r_{\rm ms}$
and then rolls over to zero at the radius of marginal stability.

The Novikov-Thorne disk model gives us a well defined starting point
for comparing theory with data.  One more step is required, however;
we need to relate the underlying dissipation in the accretion disk
(given by the Novikov-Thorne model) to the X-ray irradiation of the
disk surface (which determines the observed broadening profile of the
X-ray reflection spectrum).  For the moment, we make the simplest
assumption that the primary continuum X-ray source is located a small
distance above the disk surface (the ``local corona assumption'')
and radiates a fixed fraction of the energy dissipated in the
underlying disk.  This model has been explicitly compared with the
first {\it XMM-Newton} observation of MCG--6-30-15 which caught the
source in its enigmatic ``Deep Minimum State'' (Wilms et al. 2001;
Reynolds et al. 2004; see Iwasawa et al. 1996 for the original
identification and study of the Deep Minimum State).  Reynolds et
al. (2004) showed that the ``vanilla'' Novikov-Thorne model
supplemented by the local corona approximation {\it fails} to produce
enough relativistic broadening, even in the case of a near-extremal
($a=0.998$) rotating black hole (see Fig.~5a).  Essentially,
there is more X-ray irradiation at the smallest radii compared with
the predictions of this model.

\begin{figure*}
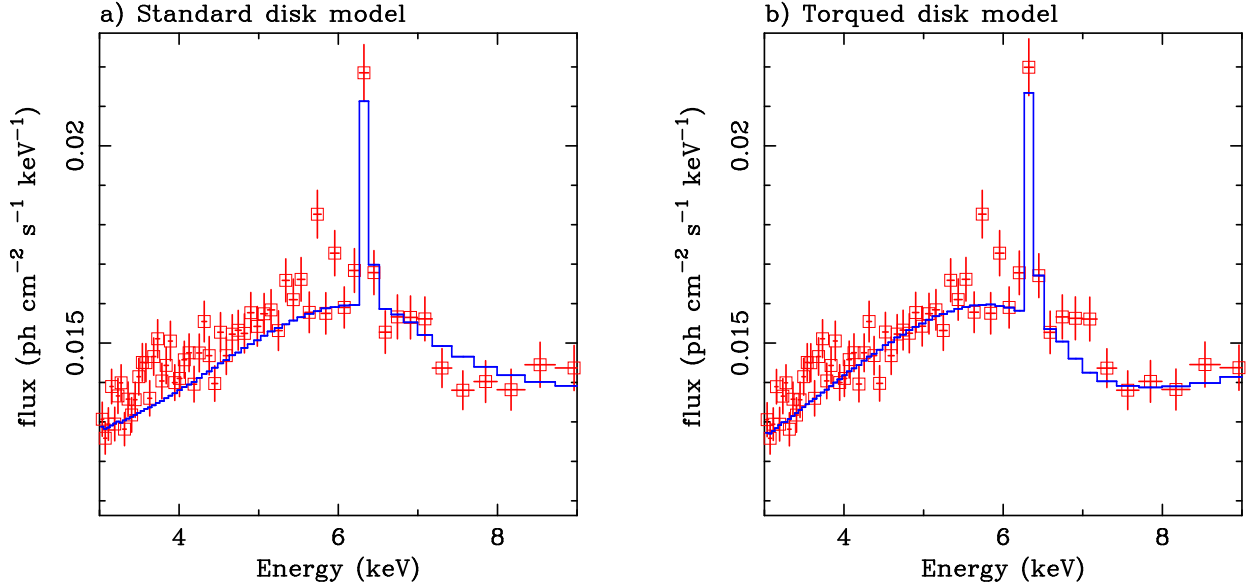

\begin{center}
\hbox{
\psfig{figure=f5a.ps,width=0.45\textwidth,angle=270}
\hspace{1cm}
\psfig{figure=f5b.ps,width=0.45\textwidth,angle=270}
}
\end{center}
\caption{Broad iron line fit assuming that the line emission tracks the 
  underlying disk dissipation of (a) a standard (Novikov \& Thorne
  1974) accretion disk and, (b) an Agol \& Krolik (2000) torqued
  accretion disk.  Modified from Reynolds et al. (2004).}
\end{figure*}

One can attempt to rescue the Novikov-Thorne disk model by supposing
that a larger portion of the total dissipation in the disk is
channeled into the X-ray emitting corona as one moves to smaller
radii.  However, since 30--50\% of the bolometric power of
MCG--6-30-15 seems to emerge through the X-ray emitting corona, one
cannot decouple it entirely from the dissipation distribution.  In the
most extreme model (which provides an adequate but not the best fit to
the data), {\it all} of the dissipated energy is channeled into the
X-ray emitting corona within the central $5GM/c^2$, while the X-ray
production efficiency is zero beyond that radius.

\subsection{Torqued accretion disks}

In attempting to reconcile the conflict between data and disk theory
noted above, one may consider either modifications to the basic
accretion disk dynamics and/or deviations from the local corona
assumption.  To begin with, we preserve the local corona assumption
and examine whether the model can be brought into line with the data
via a change of inner boundary condition.

Even when originally setting up the zero-torque boundary condition at
the radius of marginal stability, Page \& Thorne (1974) noted that
magnetic fields may allow this zero-torque boundary condition to be
violated.  Given the modern viewpoint of accretion disks, that the
very ``viscosity'' driving accretion is due to magnetohydrodynamic
(MHD) turbulence, the idea that the zero-torque boundary condition can
be violated has been revived by recent theoretical work, starting with
Gammie (1999) and Krolik (1999a).  In independent treatments, these
authors show that significant energy and angular momentum can be
extracted from matter within the radius of marginal stability via
magnetic connections with the main body of the accretion disk.  Agol
\& Krolik (2000) have performed the formal extension of the standard
Novikov-Thorne model to include a torque at $r=r_{\rm ms}$ and show
that the extra dissipation associated with this torque produces a very
centrally concentrated dissipation profile.  As shown by Gammie
(1999), Agol \& Krolik (2000), and Li (2002), this process can lead to
an extraction (and subsequent dissipation) of spin energy and angular
momentum from the rotating black hole by the accretion disk.  In these
cases, the magnetic forces might be capable of placing the innermost
part of the flow on negative energy orbits, allowing a Penrose process
to be realized (we note that Williams [2003] has also argued for the
importance of a non-magnetic, particle-particle and particle-photon
scattering mediated Penrose process).  The basic notion that the
plunging region exerts significant torques on the disk has been
verified through both pseudo-Newtonian (Hawley 2000; Reynolds \&
Armitage 2001) and fully relativistic MHD simulations of accretion
disks (e.g., McKinney \& Gammie 2004; Hirose et al. 2004).  A second
mechanism by which the central accretion disk can be torqued is via a
direct magnetic connection between the inner accretion disk and the
(rotating) event horizon of the black hole.  In this case, as long as
the angular velocity of the event horizon exceeds that of the inner
disk, energy and angular momentum of the spinning black hole can be
extracted via the Blandford-Znajek mechanism (Blandford \& Znajek
1977). We note that field lines that directly connect the rotating
event horizon with the body of the accretion disk through the plunging
region {\it are} seen in recent General Relativistic MHD simulations
of black hole accretion (e.g., Hirose et al. 2004).

Reynolds et al. (2004) has shown that a torqued disk can readily
explain the Deep Minimum spectrum provided the source is assumed to be
in a torque-dominated state (or, in the terminology of Agol \& Krolik
[2000], an ``infinite-efficiency'' state) whereby the power associated
with the innermost torque is instantaneously dominating the accretion
power (see Fig.~5b).  In other words, the X-ray data suggest that
during this Deep Minimum state of MCG--6-30-15 the power derived from
the black hole spin greatly exceeds that derived from accretion. Of
course, this state of affairs cannot last forever or else the central
black hole in MCG--6-30-15 would spin down to a point where it could
no longer provide this power.  At some point in its history, the
system must be in an accretion-dominated phase in which the black hole
is spun up.  However, even in its spin-dominated state, the spin-down
timescale of the central black hole is of the order of 100 million
years or more.  Thus we could envisage a situation in which the system
shines via a quasi-steady-state, spin-dominated accretion disk.  There
are hints, though, that accretion disks may switch between
spin-dominated and accretion-dominated on much shorter timescales.  In
its normal spectral state, the X-ray reflection features in
MCG--6-30-15 are much less centrally concentrated than in the Deep
Minimum State, suggesting that the normal state might be
accretion-dominated.  It is also important to note that this system
can switch between its normal state and the Deep Minimum State in as
little as 5--10\,ksec (Iwasawa et al.  1996), which corresponds to
only a few dynamical timescales of the inner accretion disk.  Thus it
is of interest to consider the physics of an accretion disk that
undergoes a rapid torquing event.

\begin{figure*}[t]
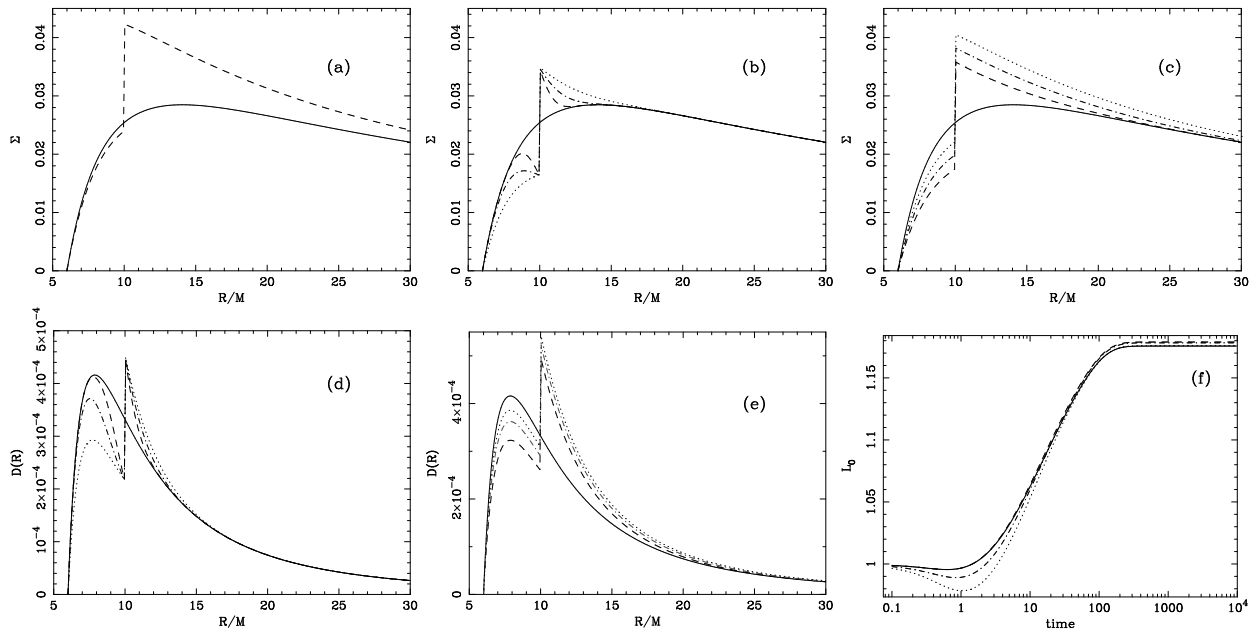

\begin{center}
\hbox{
\psfig{figure=f6a.ps,width=0.32\textwidth,angle=270}
\psfig{figure=f6b.ps,width=0.32\textwidth,angle=270}
\psfig{figure=f6c.ps,width=0.32\textwidth,angle=270}
}
\hbox{
\psfig{figure=f6d.ps,width=0.32\textwidth,angle=270}
\psfig{figure=f6e.ps,width=0.32\textwidth,angle=270}
\psfig{figure=f6f.ps,width=0.32\textwidth,angle=270}
}
\end{center}
\caption{Evolution of disk in Schwarzschild spacetime for torque at
  R/M=10.  Panel (a) shows the surface density profile just after the
  torquing event begins as well as the steady-state torqued profile is
  approached (dashed-line:$t=10000$).  Panel (b) shows the early
  stages in the evolution of the surface density profile with the
  solid line being the untorqued steady-state profile
  (dashed-line:$t=0.8$, dot-dashed-line:$t=2.53$,
  dotted-line:$t=8.0$).  Panel (c) shows the untorqued steady-state
  profile (solid-line) as well as the late-time evolution of the
  torqued profile (dashed-line:$t=25$, dot-dashed-line:$t=80$,
  dotted-line:$t=253$).  Panel (d) shows the early evolution of the
  dissipation function with lines and times corresponding to those of
  panel (b).  The qualitative feature is again a drop inward of the
  torque location and an increase outward.  Panel (e) shows the
  late-time evolution of dissipation function with lines and times
  analogous to those of panel (c) while panel (f) shows the observed
  luminosity starting at untorqued steady-state with t=0.  The
  observed luminosity is determined for angles of 10 (solid-line), 30
  (dashed-line), 60 (dot-dashed line) degrees, and 80 degrees (dotted
  line).  Although the magnitude of the observed luminosity is not the
  same in the untorqued steady-state for all angles, we have
  normalized them in order to see the change with respect to the
  untorqued state.  Note the presence of a drop in the luminosity as
  the angle of inclination decreases.  Figures from Garofalo \&
  Reynolds (2005).}
\end{figure*}

With this motivation, and guided by the time-variable inner disk
torques seen in MHD disk simulations (e.g., Reynolds \& Armitage
2001), Garofalo \& Reynolds (2005) generalized the torqued-disk models
of Agol \& Krolik (2000) to include explicit time dependence.  They
showed that the response of an initially untorqued accretion disk to a
sudden (prograde) torquing event has two phases; an initial damming of
the accretion flow together with a partial draining of the disk
interior to the torque location, followed by a replenishment of the
inner disk as the system achieves a new (torqued) steady-state.  This
is illustrated in Fig.~6 which shows the evolution of the surface
density and dissipation profiles for the example of a sporadically
torqued disk around a Schwarzschild black hole.  Garofalo \&
Reynolds (2005) propose that some of the spectral changes in
MCG--6-30-15 may be due to time-variable magnetic torques between the
disk and the plunge region or black hole itself.  In particular, they
suggest that the extremely centrally concentrated dissipation inferred
during the deep minimum state is caused by the onset of a strong inner
disk torque.  To explain why the overall X-ray luminosity of the disk
drops during this event (despite the fact that the inner torque is
doing work on the disk), Garofalo \& Reynolds (2005) suggest that the
strong returning radiation associated with this inner torque leads to
a Compton-cooling induced collapse of the X-ray emitting corona in all
but the innermost regions of the accretion disk.

\subsection{Beyond the local corona assumption: the effects of light 
bending}

Another possible explanation for a highly centrally concentrated X-ray
irradiation profile is a breakdown of the local-corona assumption.  If
the X-ray emitting source is a significant height above the
optically-thick part of the accretion disk, the hard X-ray continuum
photons will be gravitationally focused into the central regions of
the accretion disk (see Andy Fabian's contribution in these
proceedings).  Aspects of this scenario have been explored by several
authors including Martocchia \& Matt (1996), Reynolds \& Begelman
(1997) and Miniutti \& Fabian (2004).

This suggests an alternative picture for MCG--6-30-15 in which the
Deep Minimum State is produced when the X-ray source is located at
mid/high latitudes very close to the black hole.  Reynolds \& Begelman
(1997) showed that both the changes in the iron line profiles (as then
seen by {\it ASCA}) {\it and} the change in the level of the primary
continuum can be reproduced by a model in which a fairly constant
compact X-ray source on the symmetry/spin axis of the accretion disk
is shifting its vertical location.  When the source is close to the
black hole, the primary X-rays are strongly focused onto the central
disk leading to both a very broad line and a diminished observed
primary X-ray continuum flux.  Miniutti \& Fabian (2004) have extended
this model and shown that it provides an excellent description of
spectral variability seen in all phases of the {\it XMM-Newton} data
for MCG--6-30-15.

We note that the light bending scenario does not diminish the need for
exotic spin-related astrophysics --- the base of a spin-driven
magnetic jet is an obvious candidate for this elevated continuum X-ray
source.

\subsection{The mysterious variability of the iron line and the 
Compton reflection hump}

While both complex accretion disk dynamics (e.g., magnetic torques
between the plunging region and the disk-proper) and light bending of
the primary X-ray emission almost certainly are both relevant, it is
useful to ask how we might distinguish between these two straw-man
models.  Spectral variability is likely the key.

\begin{figure*}
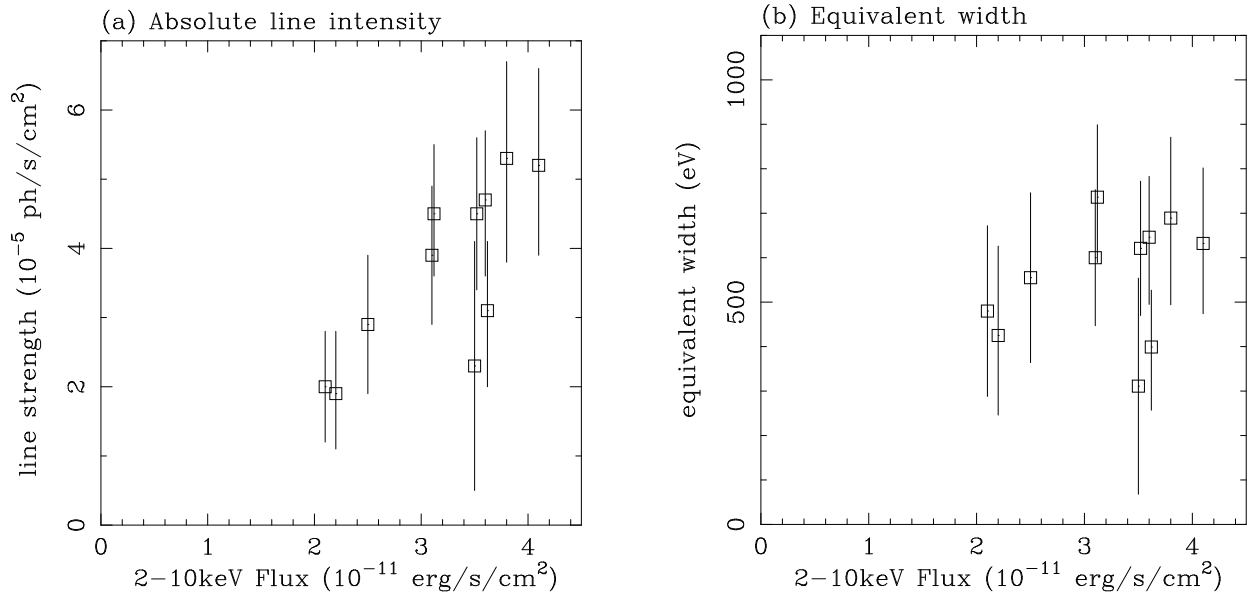

\begin{center}
\hbox{
\psfig{figure=f7a.ps,width=0.45\textwidth,angle=270}
\hspace{1cm}
\psfig{figure=f7b.ps,width=0.45\textwidth,angle=270}
}
\end{center}
\caption{Result of fitting a simple absorbed power-law plus broad iron
line component to the 10\,ksec segments of data.  Panel (a) shows the
absolute intensity of the broad line component as a function of
2--10\,keV flux.  Note the apparent correlation of line intensity with
continuum flux.  Panel (b) shows the equivalent width of the {\tt
laor} component as a function of 2--10\,keV flux.  It can be seen that
these data are consistent with a constant equivalent width.  Error
bars are shown at the 90\% level for one significant parameter
($\Delta\chi^2=2.71$). Figures from Reynolds et al. (2004).}
\end{figure*}

Naively, if the geometry of the system and the ionization state of the
disk remain unchanged, one expects the fluorescent iron line flux to
track the strength of the primary X-ray continuum (i.e., one expected
a constant iron line equivalent width).  While this is seen in {\it
XMM-Newton} data of MCG--6-30-15 for the low flux states (Reynolds et
al. 2004; also see Fig.~7), the iron line {\it flux} appears to
``saturate'' at higher fluxes implying that the equivalent width
decreases with increasing continuum flux (Fabian et al. 2002).  As
shown by Miniutti \& Fabian (2004), this behaviour has a natural
explanation within the light-bending scenario, with the strong light
bending changing the effective geometry of the system in just the
correct manner.  Within the local-corona, torqued-disk scenario, one
needs to posit changes in the ionization state of the surface layers
of the disk which conspire to cause an effective saturation of the
total line flux.

An equally important but little discussed mystery comes from examining
the dependence between the iron line flux and the strength of the
Compton reflection hump (seen at 20--30\,keV).  Naively, assuming a
fixed ionization state of the reflector, the iron line equivalent
width should be proportional to the relative flux associated with the
Compton hump (compared with the direct continuum flux) since both of
these spectral features are produced by the X-ray reflection process.
The spectral variability in the Miniutti \& Fabian (2004) light
bending model is primarily due to a change in the fraction of primary
X-ray continuum photons are directed into the disk plane and, hence,
would predict that this proportionality be preserved.  This is not
what is observed.  Observations by the Proportional Counter Array
(PCA) on the {\it Rossi X-ray Timing Explorer (RXTE)} allows us to
measure both the iron line equivalent width and the reflection hump.
In a phenomenon first found by Chiang et al. (2000) in NGC~5548, Lee
et al. (2001) showed that the iron line equivalent width and the
relative normalization of the Compton reflection hump are {\it
anti-correlated}.  This result remains valid (and is more generally
true for Seyfert 1 nuclei as a class) when the {\it RXTE}/PCA data are
re-analyzed using the latest refinements of the background model and
spectral calibration (Mattson, Weaver \& Reynolds, in preparation).

If this behaviour is further confirmed (e.g., by {\it Suzaku}), it
will strongly suggest that changes in the ionization state of the
accretion disk are playing an important part in determining the
spectral variability and must be explicitly modeled.

\section{Conclusions and future hopes}
\label{sec: conclusions}

Current data from {\it XMM-Newton} and {\it Chandra} are already
allowing us to probe black hole physics within a few gravitational
radii of the event horizon, and may well be giving us the first
observational glimpses of physics within the ergosphere.  But this is
just the beginning of X-ray astronomy's exploration of strong gravity
and black hole accretion, not the end of the road.  The enormous
throughput of {\it Constellation-X} will allow us to probe detailed
time variability of the iron line.  Dynamical timescale line
variability, an easy goal for {\it Constellation-X}, will allow us to
follow non-axisymmetric structures in the disk as they orbit (Armitage
\& Reynolds 2004; also see Iwasawa, Miniutti \& Fabian [2004] for the
first hint of such structure in {\it XMM-Newton} data).  This gives us
a direct probe of an almost Keplerian orbit close into a black hole.
Furthermore, line variability on the light crossing time will allow us
to probe relativistic reverberation signatures (Reynolds et al. 1999;
Young \& Reynolds 2000), essentially giving us a direct probe of the
null geodesics in the space-time.  Together, these variability
signatures will take their place along side gravitational wave
experiments in allowing true tests of strong-field GR.

There is no compelling reason to believe that GR fails on the
macroscopic scales probed by either X-ray or gravitational wave
studies of astrophysical black holes.  In the event that GR is
verified, both X-ray and gravitational wave observations will allow
unambiguous measurements of black hole spins.  Gravitational wave
observations with {\it LISA} of a stellar mass black hole spiraling
into a $10^6\,M_\odot$ black hole (a so-called Extreme Mass Ratio
Inspiral; EMRI) will allow precision measurement of the supermassive
black hole's spin as well as tests of the no-hair theorem and the Kerr
metric.  X-ray spectroscopy with {\it Constellation-X} provides a
crucial parallel track of study in which we can obtain measurements of
black hole spin across the whole mass range of astrophysical black
holes (i.e., stellar, intermediate, and supermassive) using spectral
features that are already known to exist.  Only then can the
demographics and astrophysical relevance of black hole spin truly be
gauged.

\acknowledgements

We thank the conference organizers for a stimulating meeting.  We also
thank support from the National Science Foundation under grant
AST0205990.

\end{document}